\documentclass[aps,superscriptaddress,nofootinbib,floatfix,notitlepage]{revtex4-1}
\usepackage[utf8x]{inputenc}
\pdfoutput=1
\usepackage{graphicx}
\usepackage{hyperref}
\usepackage{bm}
\usepackage[sort&compress]{natbib}

\begin{document}

\title{Testing the Standard Model and searching for New Physics with
  $B_d \to \pi \pi$ and $B_s \to K K$ decays}

\author{M.~Ciuchini}
\affiliation{INFN, Sezione di Roma Tre, Via della Vasca Navale 84, I-00146 Roma, Italy}
\author{E.~Franco}
\affiliation{INFN, Sezione di Roma, Piazzale A. Moro 2, I-00185 Roma, Italy}
\author{S.~Mishima}
\affiliation{INFN, Sezione di Roma, Piazzale A. Moro 2, I-00185 Roma, Italy}
\author{L.~Silvestrini}
\affiliation{INFN, Sezione di Roma, Piazzale A. Moro 2, I-00185 Roma, Italy}

\begin{abstract}
  We propose to perform a combined analysis of $B \to \pi\pi$ and $B_s
  \to K^+ K^-$ modes, in the framework of a global CKM fit. The method
  optimizes the constraining power of these decays and allows to
  derive constraints on NP contributions to penguin amplitudes or on
  the $B_s$ mixing phase. We illustrate these capabilities with a
  simplified analysis using the recent measurements by the LHCb
  Collaboration, neglecting correlations with other SM observables.
\end{abstract}
\maketitle

CP violation in $B_{d,s}$ decays plays a fundamental role in testing
the consistency of the Cabibbo-Kobayashi-Maskawa (CKM) paradigm in
the Standard Model (SM) and in probing virtual effects of heavy new particles.
With the advent of the B-factories, the Gronau-London (GL) isospin
analysis of $B_d \to \pi \pi$ decays \cite{Gronau:1990ka} has been a
precious source of information on the phase of the CKM matrix.
Although the method allows a full determination of the weak phase and of
the relevant hadronic parameters, it suffers from discrete ambiguities that
limit its constraining power. It is however possible to reduce the impact of
discrete ambiguities by adding information on hadronic
parameters \cite{Charles:1998qx,Bona:2007qta}. In particular, as noted
in refs.~\cite{Fleischer:1999pa,Fleischer:2007hj,Fleischer:2010ib}, the hadronic
parameters entering the $B_d \to \pi^+ \pi^-$ and the $B_s \to K^+
K^-$ decays are connected by U-spin, so that the experimental
knowledge of $B_s \to K^+ K^-$ can definitely improve the extraction
of the CKM phase with the GL analysis. Indeed, in
ref.~\cite{Bona:2007qta}, the measurement of $\mathrm{BR}(B_s \to K^+
K^-)$ was used to obtain an upper bound on one of the hadronic
parameters. 

After the pioneering studies performed at the TeVatron, very recently
LHCb opened up the road to CP violation in $B_s \to K K$ decays
\cite{LHCbconf}. The present experimental information is summarized in
Table \ref{tab:exp}. At present, one has all the necessary information
to use the U-spin strategy proposed by Fleischer (F) in
refs.~\cite{Fleischer:1999pa,Fleischer:2007hj} to extract the CKM phase
from a combined analysis of $B_d \to \pi^+ \pi^-$ and the $B_s \to K^+
K^-$ decays. However, as we will show explicitly below, this strategy
alone suffers from a sizable dependence on the breaking of U-spin
symmetry \cite{Beneke:2003zw}.

Furthermore, in the $B_s$ system the measurement of any time-dependent
CP asymmetry cannot be directly translated into a measurement of the
angle $\beta_s =
\mathrm{arg}\,\left(-\frac{V_{tb}^*V_{ts}}{V_{cb}^*V_{cs}}\right)$,
even in the case of the so-called ``gold-plated'' $b \to c \bar c s$
decays. This is due to the fact that the angle $\beta_s$ is small and
correlated to the subdominant amplitude in $b \to c \bar c s$
decays. Thus, measuring $\beta_s$ requires the determination of the
subdominant decay amplitude. This is evident by noting that using CKM
unitarity the $b \to s$ decay amplitude can be written as
\begin{equation}
  \label{eq:pars}
  A = V_{ub}V_{us}^* T + V_{cb}V_{cs}^* P \quad \mathrm{or} \quad A =
  V_{ub}V_{us}^* (T - P) + V_{tb}V_{ts}^* (-P). 
\end{equation}
Naively dropping the doubly Cabibbo-suppressed term proportional to
$V_{ub}V_{us}^*$ would lead to the conclusion that the CP asymmetry
measures $2\beta_s$ with the first choice or that the CP asymmetry
should vanish in the second choice. Clearly, a full treatment of the
decay amplitude, taking into account correlations between the various
CKM terms, is necessary to give a meaningful interpretation to the CP
asymmetry. This is at variance with the $B_d$ case, where the angle
$\beta$ is large and thus the time-dependent CP asymmetry in $b \to c
\bar c s$ decays gives sin$2\beta$ with a good accuracy.\footnote{The
  uncertainty due to the subleading amplitude can be quantified using
  $B_d \to J/\psi \pi$, and the correlation with the CKM terms in the
  decay amplitude is negligible
  \cite{Ciuchini:2005mg,Faller:2008zc,Gronau:2008cc,Ciuchini:2011kd}} 
In this respect, the combined analysis of the GL modes and $B_s \to
K^+ K^-$ is optimal, since one has full knowledge of the U-spin related
control channel $B_d \to \pi^+ \pi^-$, similarly to the case of $B_s
\to K^{(*)0} \bar K^{(*)0}$ and $B_d
\to K^{(*)0} \bar K^{(*)0}$ proposed in
ref.~\cite{Ciuchini:2007hx}. Conversely, the ``gold-plated'' $B_s \to
J/\psi \phi$ decay has no U-spin related control channel, making the
extraction of $\beta_s$ problematic \cite{Faller:2008gt}.

\begin{table}
 \begin{tabular}{|c| c| c| c| c| c|}
  \hline
  Channel & BR $\times 10^6$ & $\mathcal{S} (\%)$ & $\mathcal{A}_\mathrm{CP}(=-\mathcal{C}) (\%)$ & corr. & ref. \\
  \hline
  $B_d \to \pi^+ \pi^-$ & $5.11 \pm 0.22$ & $-65 \pm 7$ & $38 \pm 6 $
  & $0.08$ &
\cite{Aubert:2008sb,Ishino:2006if,Aubert:2006fha,belle,Bornheim:2003bv,Aaltonen:2011qt} \\
  $B_d \to \pi^+ \pi^-$ & -- & $-56 \pm 17 \pm 3$ & $11 \pm 21 \pm 3$
  & $-0.34$ & \cite{LHCbconf} \\
  $B_d \to \pi^0 \pi^0$ & $1.91 \pm 0.23$ & -- & $43 \pm 24$ & -- &
\cite{Aubert:2008sb,Abe:2004mp,Bornheim:2003bv} \\
  $B^+ \to \pi^+ \pi^0$ & $5.48 \pm 0.35$ & -- & $2.6 \pm 3.9$ & -- &
\cite{Aubert:2007hh,belle,Bornheim:2003bv} \\
  $B_s \to K^+ K^-$ & $25.4 \pm 3.7$ & $17 \pm 18 \pm 5$ & $2 \pm 18
  \pm 4$ & $-0.1$ & \cite{Peng:2010ze,Aaltonen:2011qt,LHCbconf}
\\
\hline
\end{tabular}
\caption{Experimental data used in the analysis. The correlation column refers
to the $\mathcal{S}$ and $\mathcal{A}_\mathrm{CP}$ measurements. Except from the results in
ref.~\cite{LHCbconf}, all other measurements have been averaged by
HFAG \cite{*[][{ and online update at
    http://www.slac.stanford.edu/xorg/hfag}] HFAG}. The CP asymmetry of $B^+ \to \pi^+ \pi^0$ has been reported
for completeness, although it has not been used in the analysis.}
\label{tab:exp}
\end{table} 

We propose to perform a combined analysis of the GL modes plus $B_s
\to K^+ K^-$, including the time-dependent CP asymmetries, to obtain
an optimal determination of the CKM phase within the SM. We show that
this combined strategy has a mild dependence on the magnitude of
U-spin breaking, allowing for a solid estimate of the theory
error.

Beyond the SM, NP can affect both the $B_{d,s} - \bar B_{d,s}$
amplitudes and the $b \to d, s$ penguin amplitudes. Taking the phase
of the mixing amplitudes from other measurements, for example from $b
\to c \bar c s$ decays, one can obtain a constraint on NP in $b \to s$
penguins. Alternatively, assuming no NP in the penguin amplitudes, one
can obtain a constraint on NP in mixing.

In this letter, we illustrate the points above in a simplified
framework, neglecting SM correlations with other observables and using
as input values sin$2 \beta = 0.679 \pm 0.024$ \cite{HFAG} and $2
\beta_s =(0 \pm 5)^\circ$ \cite{LHCb-CONF-2012-002}, obtained from $b
\to c \bar c s$ decays.  Clearly, the optimal strategy will be to
include the combined analysis of the GL and F modes in a global fit of
the CKM matrix plus possible NP contributions.

The GL and F analyses were formulated with different parameterizations
of the decay amplitudes. In order to use the constraints in a global
fit, one should write the decay amplitudes with the full dependence on
CKM matrix elements, but for the present analysis we can choose the F
one and write the amplitudes as follows:
\begin{equation}
  \label{eq:ampli}
  \begin{array}{ll}
  A(B_d \to \pi^+ \pi^-) = C  (e^{i \gamma} - d e^{i \theta})\,, &
  A(\bar B_d \to \pi^+ \pi^-) = C  (e^{-i \gamma} - d e^{i
    \theta})\,, \\
  A(B_d \to \pi^0 \pi^0) = \frac{C}{\sqrt{2}}  (T e^{i \theta_T} e^{i \gamma} + d e^{i \theta})\,, &
  A(\bar B_d \to \pi^0 \pi^0) = \frac{C}{\sqrt{2}}  (T e^{i \theta_T} e^{-i \gamma} + d e^{i \theta})\,,  \\
  A(B^+ \to \pi^+ \pi^0) = \frac{A(B_d \to \pi^+ \pi^-)}{\sqrt{2}} + A(B_d \to \pi^0 \pi^0)\,, &
  A(B^- \to \pi^- \pi^0) = \frac{A(\bar B_d \to \pi^+ \pi^-)}{\sqrt{2}} +
  A(\bar B_d \to \pi^0 \pi^0) \,,\\
  A(B_s \to K^+ K^-) = C^\prime \frac{\lambda}{1-\lambda^2/2} 
  (e^{i \gamma} + \frac{1-\lambda^2}{\lambda^2} d^\prime e^{i
    \theta^\prime})\,, &
  A(\bar B_s \to K^+ K^-) = C^\prime \frac{\lambda}{1-\lambda^2/2}
   (e^{-i \gamma} + \frac{1-\lambda^2}{\lambda^2} d^\prime e^{i \theta^\prime})\,,
 \end{array}
\end{equation}
where the magnitude of $V_{ub}V_{ud}$ has been reabsorbed in $C$, 
the magnitude of $V_{cb}V_{cd}/(V_{ub}V_{ud})$ has been reabsorbed in
$d$ and $\lambda = 0.2252$.
In the exact U-spin limit, one has $C = C^\prime$, $d = d^\prime$ and
$\theta = \theta^\prime$. We have neglected isospin breaking in $B_d
\to \pi \pi$, since its impact on the extraction of the weak phase is
at the level of
$1^\circ$~\cite{Gronau:1998fn,Zupan:2007fq,Gardner:2005pq,Botella:2006zi}. 
The physical observables entering the analysis are:
\begin{eqnarray}
  \label{eq:obs}
  &&BR(B \to MM) = F(B) \frac{\vert A(B \to MM) \vert^2 + \vert A(\bar B \to
    MM) \vert^2}{2}\,,  \\ &&\mathcal{A}_\mathrm{CP} = - \mathcal{C}
  = \frac{\vert A(\bar B \to MM) \vert^2 - \vert A(B \to
    MM) \vert^2}{\vert A(\bar B \to MM) \vert^2 + \vert A(B \to
    MM) \vert^2}\,, \qquad
  \mathcal{S} = \frac{2 \mathrm{Im}
  \left(e^{-i \phi_M(B)}
  \frac{A(\bar B \to MM)}{A(B \to
    MM)}
  \right)}{1 +
  \left\vert  \frac{A(\bar B \to MM)}{A(B \to
    MM)} \right\vert^2}\,,\nonumber
\end{eqnarray}
where $F(B_d) = 1$, $F(B^+) = \tau_{B^+}/\tau_{B_d} = 1.08$, $F(B_s) =
\Phi(B_s) (2-(1-y_s^2)\tau(B_s \to K^+ K^-)/\tau_{B_s} )$
\cite{deBruyn:2012wj}, $\tau_{B_s} = (1.425 \pm 0.041) \mathrm{ps}$, 
$\Phi(B_s) = \tau_{B_s}/\tau_{B_d} (m_{B_d}^2/m_{B_s}^2)
\sqrt{(M_{B_s}^2-4 M_{K^+}^2)/(M_{B_d}^2-4 M_{\pi^+}^2)} = .9112$,
$y_s = \Delta \Gamma_s/(2\Gamma_s) = (0.149 \pm 0.015)/2$
\cite{Bona:2007vi}, $\tau(B_s \to K^+ K^-) = (1.463 \pm 0.042) $ ps
\cite{Aaij:2012kn,LHCb-CONF-2012-001} and $\phi_M(B_d) = 2 \beta$,
$\phi_M(B_s) = -2 \beta_s$ in the SM.

\begin{figure}[t]
  \centering
  \includegraphics[width=.23\textwidth]{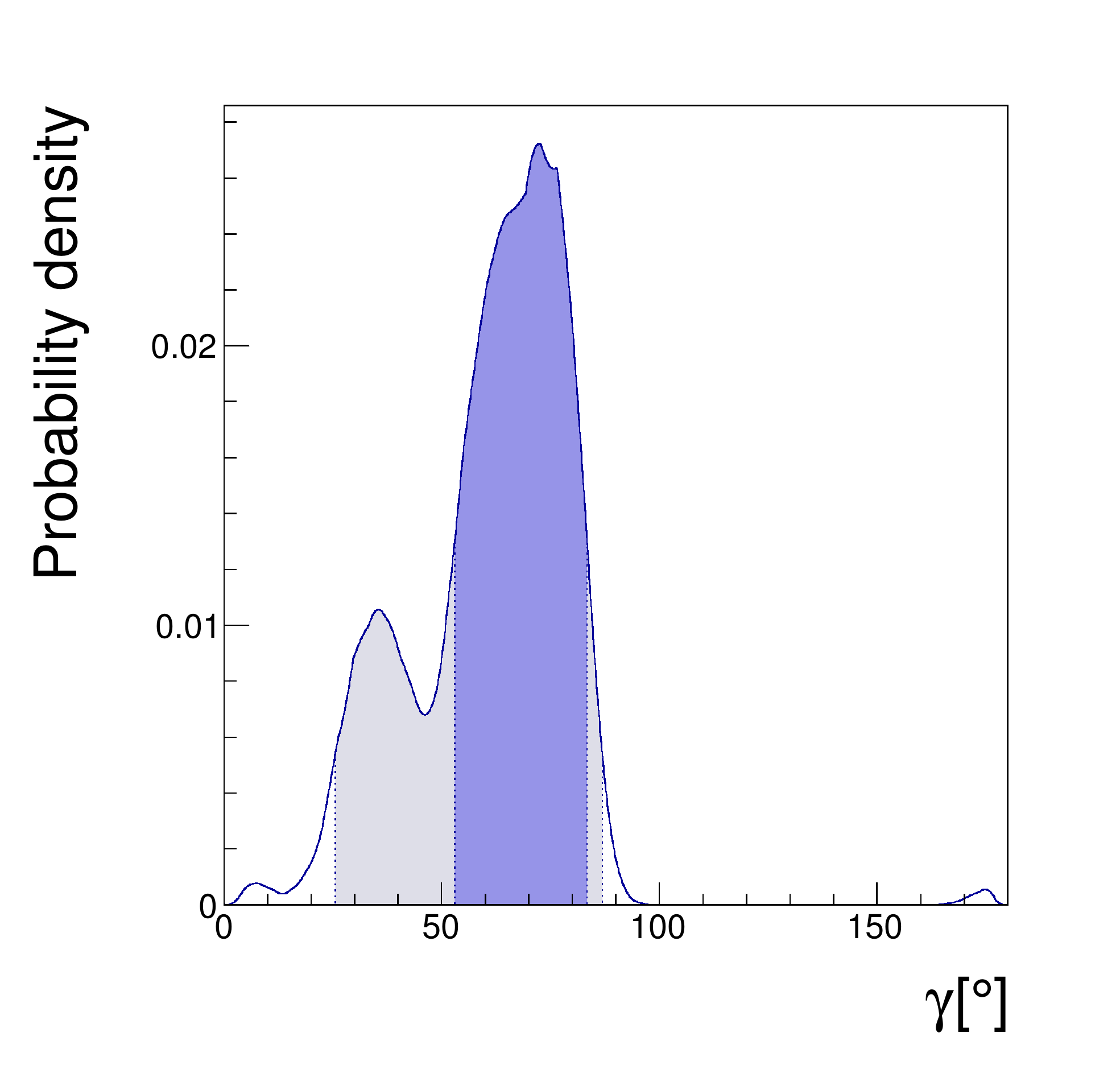}
  \includegraphics[width=.23\textwidth]{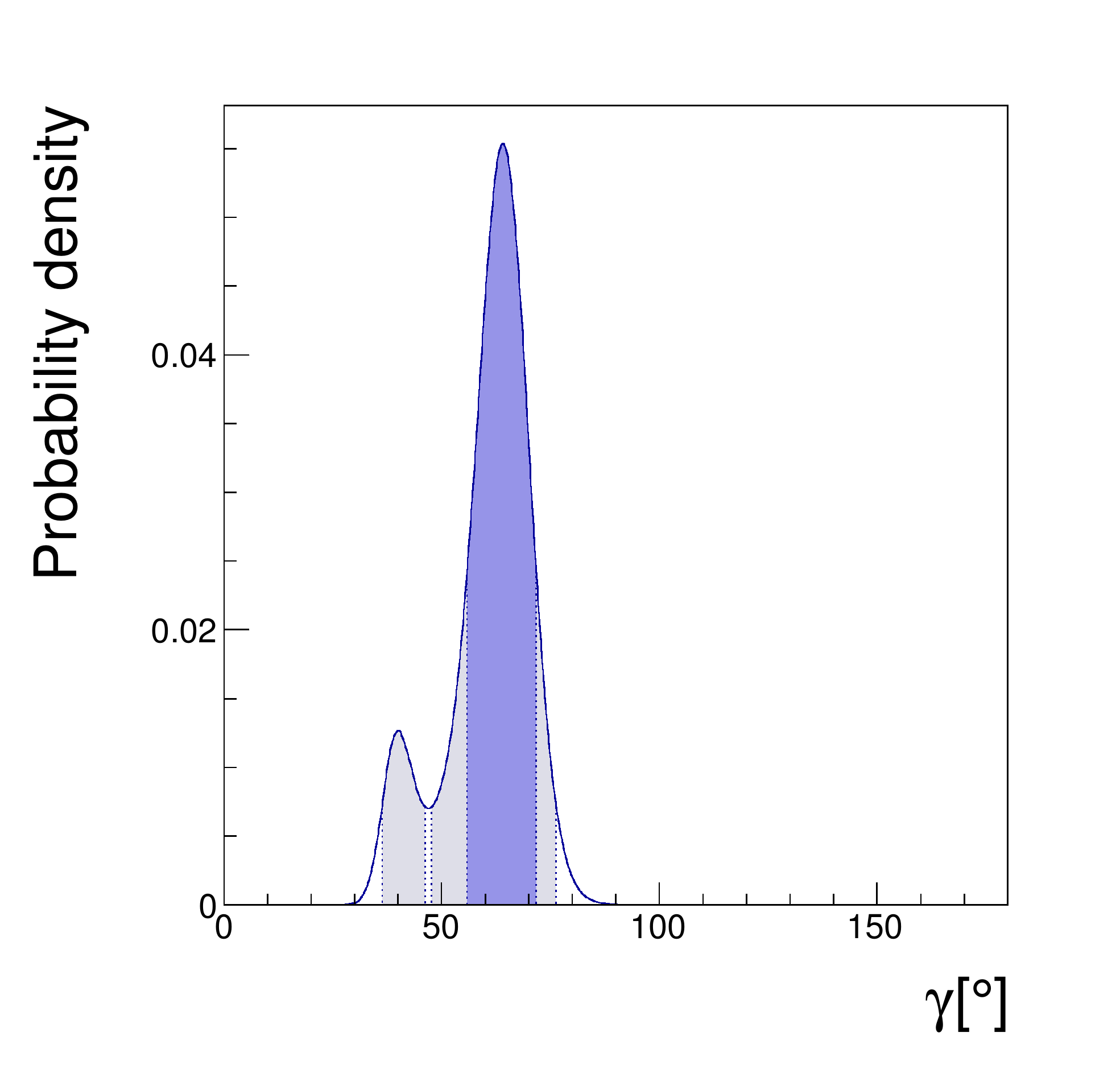}
  \includegraphics[width=.23\textwidth]{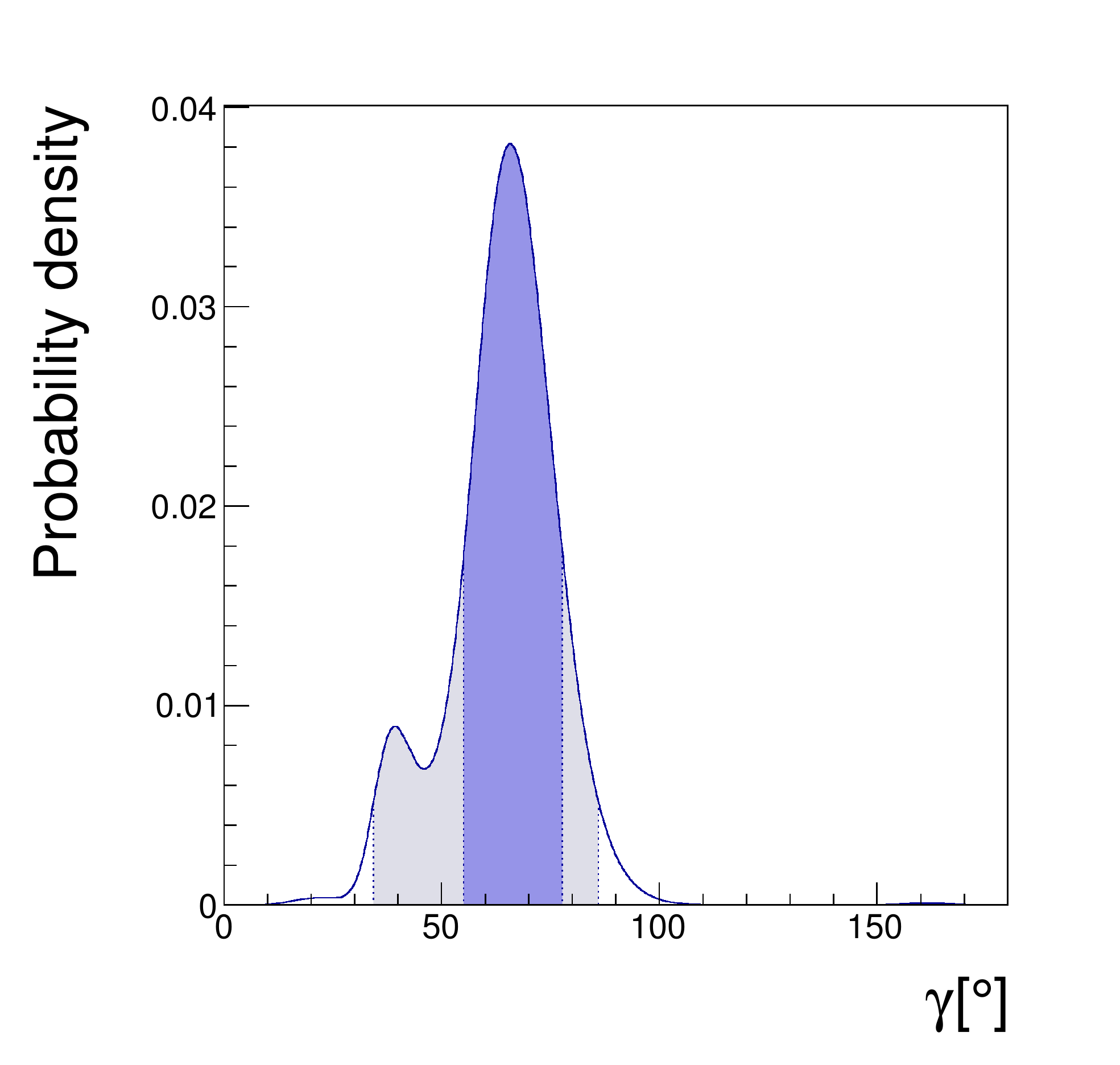}
  \includegraphics[width=.23\textwidth]{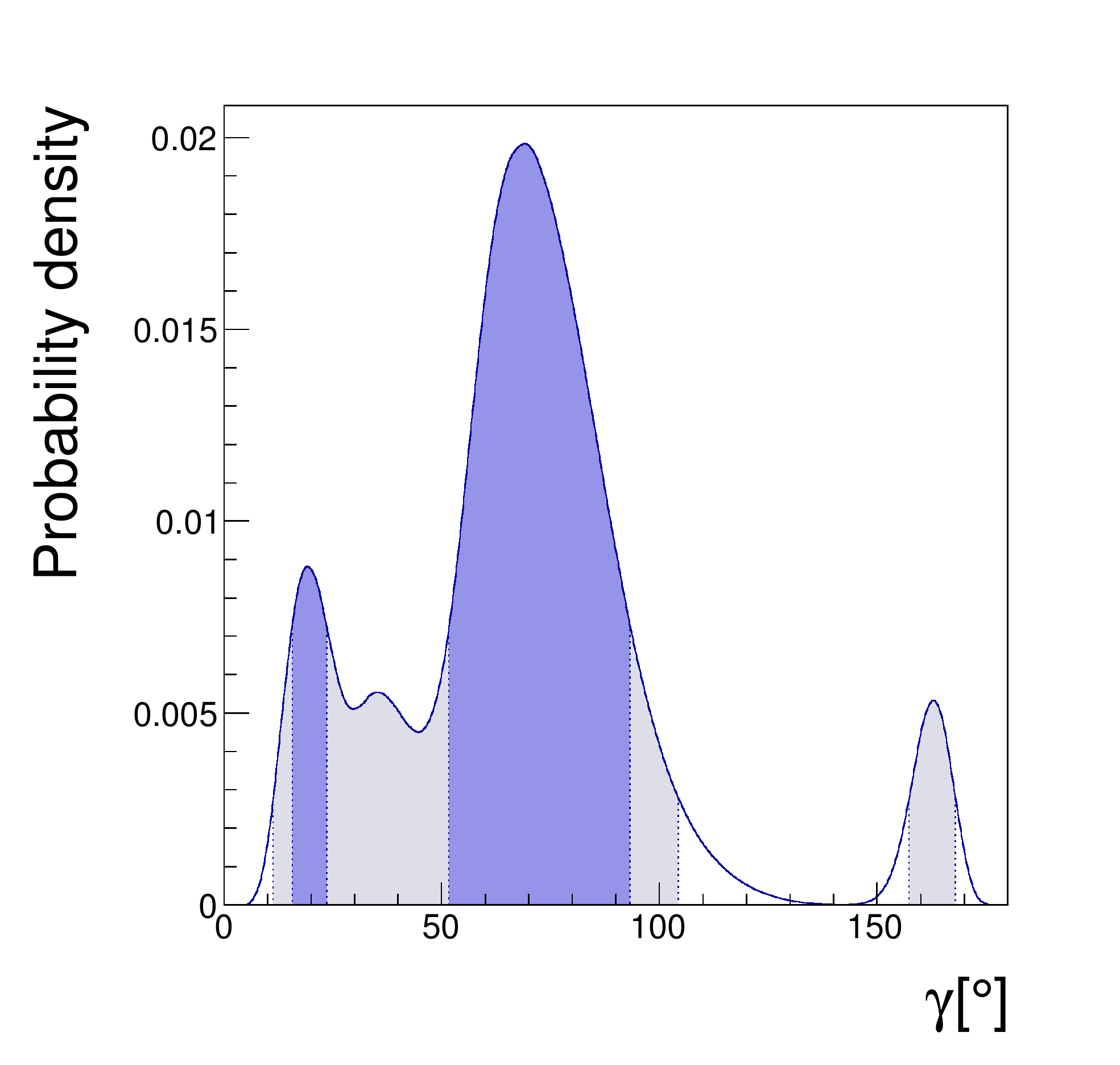}
  \caption{From left to right: P.d.f. for $\gamma$ obtained using the
    GL method as described in the text; p.d.f. for $\gamma$ obtained
    using the F method for $\kappa = 0.1$, $0.3$, $0.5$. Here and in
    the following, dark (light) areas correspond to $68\%$ ($95\%$)
    probability regions.}
  \label{fig:gammaGL}
\end{figure}

In the GL approach, one extracts the p.d.f. for the angle $\alpha =
\pi - \beta - \gamma$ of the Unitarity Triangle (UT) from the
measurements of the three $BR(B\to \pi\pi)$, $\mathcal{S}(\pi^+
\pi^-)$, $\mathcal{A}_\mathrm{CP}(\pi^+ \pi^-)$ and
$\mathcal{A}_\mathrm{CP}(\pi^0 \pi^0)$.\footnote{Using unitarity of
  the CKM matrix, it is possible to write the $B \to \pi \pi$ decay
  amplitudes and observables in terms of $\alpha$ instead of $\gamma$
  and $\beta$. However, for the purpose of connecting $B \to \pi \pi$
  to $B_s \to K K$ it is more convenient to use the parameterization
  in eq.~(\ref{eq:ampli}).} In this way, $\alpha$ (or, equivalently,
$\gamma$), is determined up to discrete ambiguities, that correspond
however to different values of the hadronic parameters. As discussed
in detail in ref.~\cite{Bona:2007qta}, the shape of the
p.d.f. obtained in a Bayesian analysis depends on the allowed range
for the hadronic parameters. For example, using the data in Table
\ref{tab:exp}, solving for $C$ and choosing flat a-priori
distributions for $d\in[0,2]$, $\theta\in[-\pi,\pi]$,
$T\in[0,1.5]$ and $\theta_T\in[-\pi,\pi]$ we obtain the p.d.f. for
$\gamma$ in Fig.~\ref{fig:gammaGL}, corresponding to $\gamma = (68 \pm
15)^\circ$ ($\gamma \in [25, 87]^\circ$ at $95 \%$ probability). Here
and in the following we plot $\gamma$ only in the range
$[0,180]^\circ$ since the result is periodic with period $180^\circ$.  

Using instead the F method, one can obtain a p.d.f. for $\gamma$ from
$BR(B\to \pi^+ \pi^-)$, $BR(B_s\to K^+ K^-)$, $\mathcal{S}(\pi^+
\pi^-)$, $\mathcal{A}_\mathrm{CP}(\pi^+ \pi^-)$, $\mathcal{S}(K^+
K^-)$ and $\mathcal{A}_\mathrm{CP}(K^+ K^-)$ given
a range for the U-spin breaking effects. Fleischer suggested to
parameterize the U-spin breaking in $C^\prime/C$ using the result one
would obtain in factorization, namely
\begin{equation}
  \label{eq:rfact}
  r_\mathrm{fact} = \left\vert
    C^\prime/C\right\vert_\mathrm{fact} = 1.46 \pm 0.15\,,
\end{equation}
where we have symmetrized the error obtained using light-cone QCD sum
rule calculations in ref.~\cite{Duplancic:2008tk}. However, this can
only serve as a reference value, since there are nonfactorizable
contributions to $C$ and $C^\prime$ that could affect this estimate
\cite{Beneke:2003zw}. In our analysis, we parametrize nonfactorizable
U-spin breaking as follows:
\begin{equation}
  \label{eq:su3b}
  C^\prime = r_\mathrm{fact} r_C C\,,
  \qquad
  d^\prime e^{i \theta^\prime} = d e^{i
    \theta} + r_d d e^{i r_\theta}\,,
\end{equation}
with $r_C$, $r_d$ and $r_\theta$ uniformly distributed in the range
$[1-\kappa, 1+\kappa]$, $[0,\kappa]$ and $[-\pi,\pi]$ respectively.

In Fig.~\ref{fig:gammaGL} we present the p.d.f. for $\gamma$ obtained
with the F method for three different values of $\kappa$. We see that
the method is very precise for $\kappa = 0.1$, it is comparable to the
GL method for $\kappa = 0.3$, and it becomes definitely worse for
$\kappa = 0.5$. Thus, a determination of $\gamma$ from the F method
alone is subject to the uncertainty on the size of U-spin breaking. 

\begin{figure}[t]
  \centering
  \includegraphics[width=.23\textwidth]{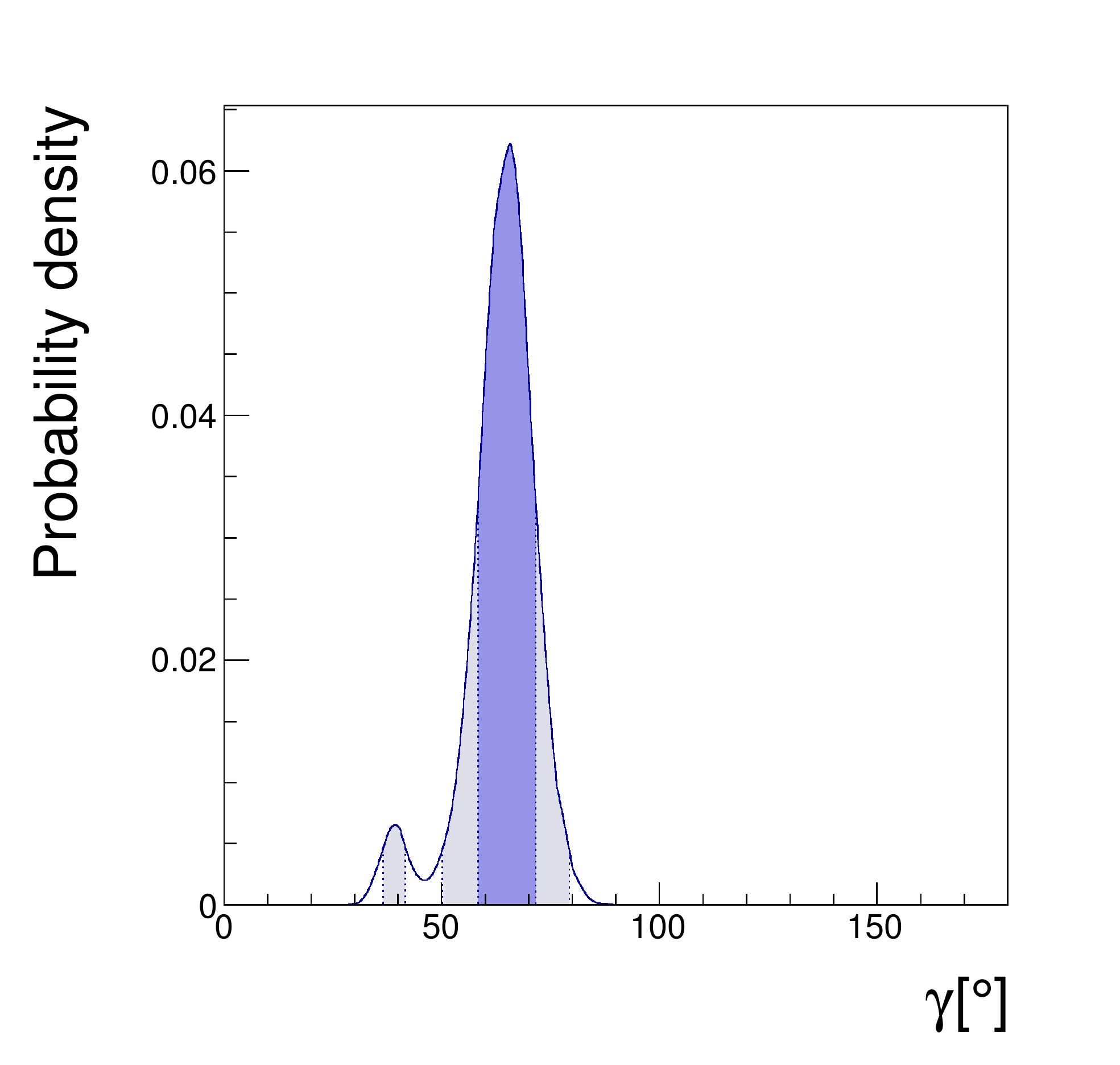}
  \includegraphics[width=.23\textwidth]{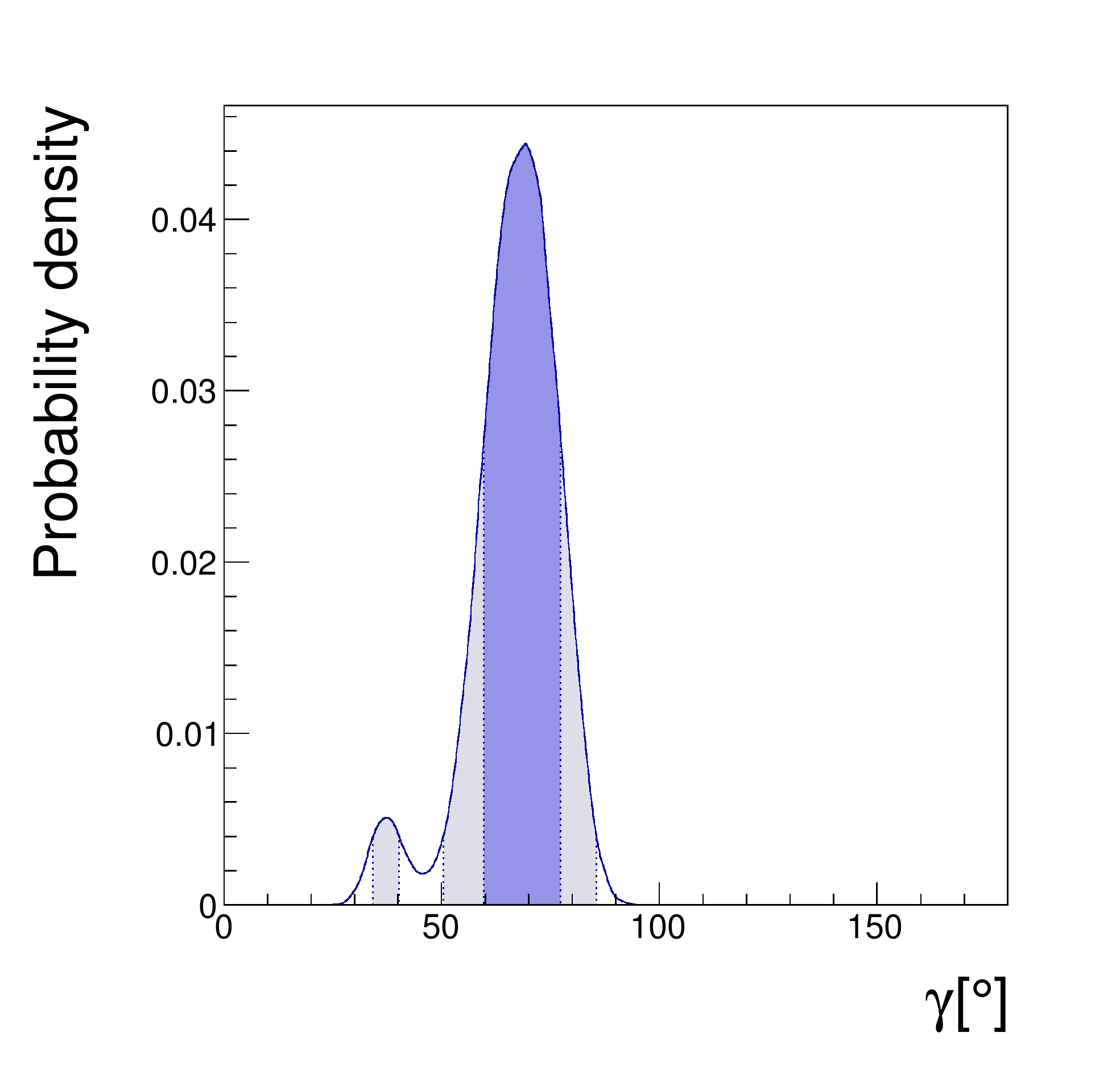}
  \includegraphics[width=.23\textwidth]{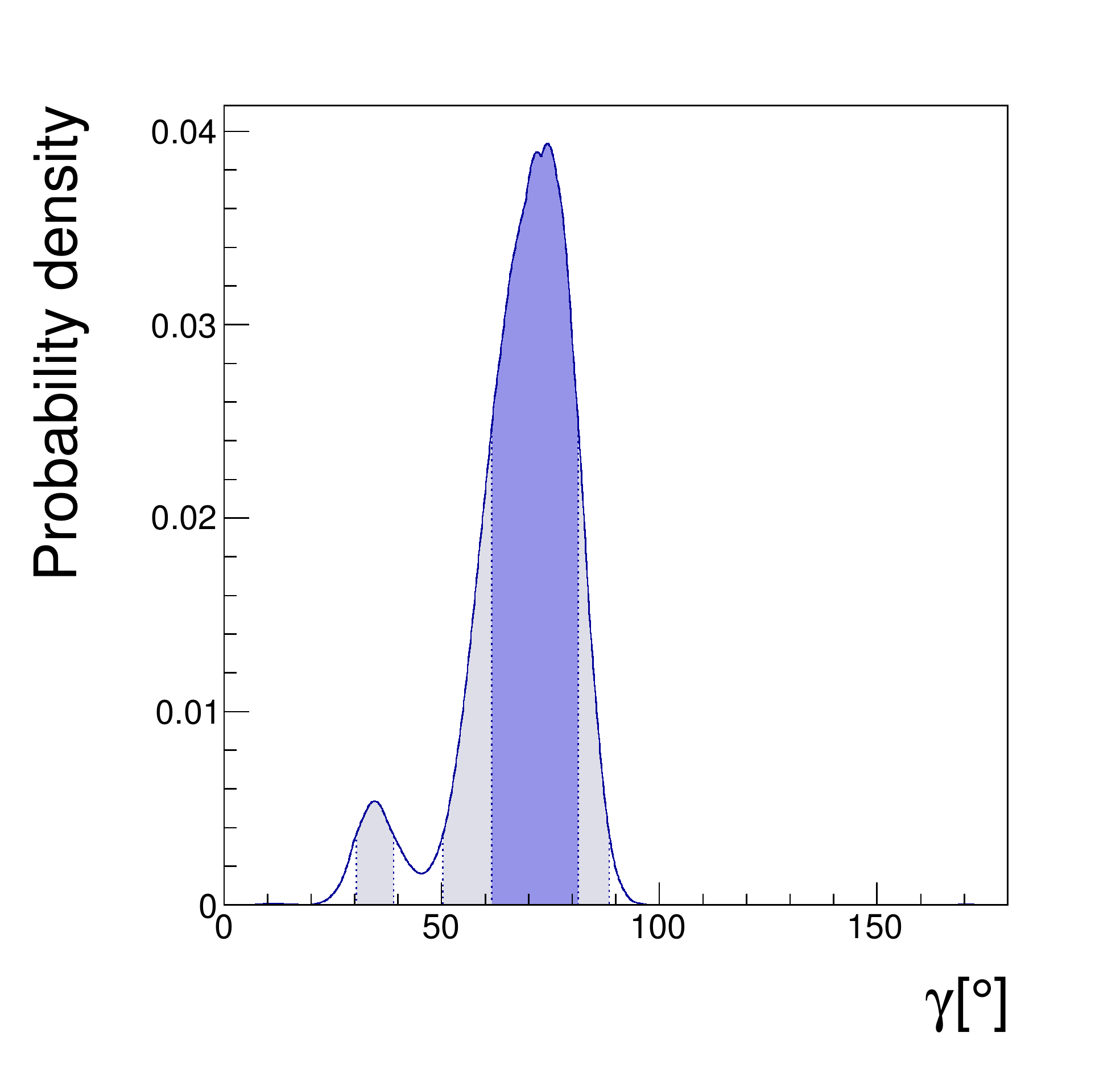}
  \includegraphics[width=.23\textwidth]{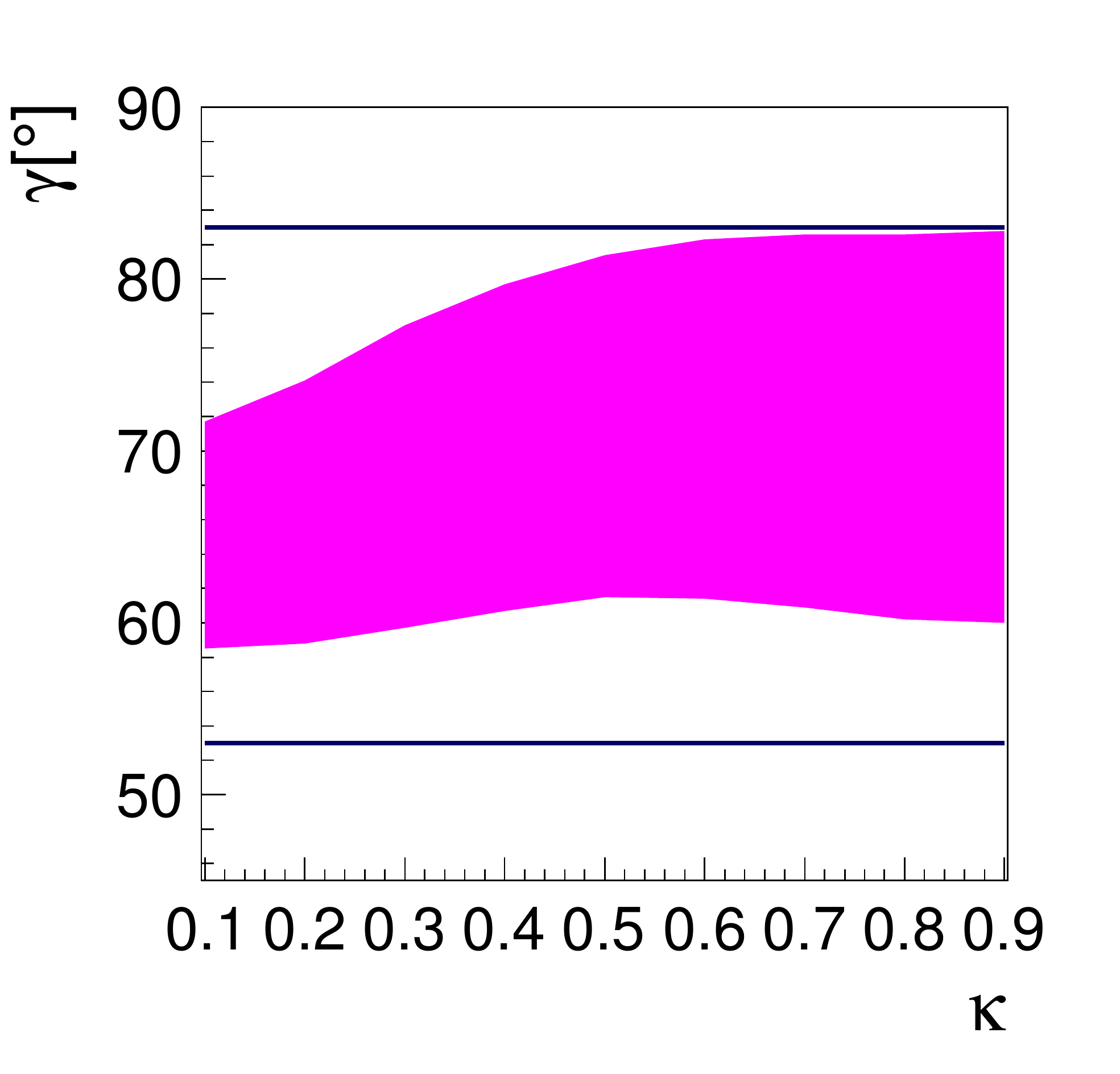}
  \caption{From left to right: P.d.f. for $\gamma$ obtained using the
    combined method for $\kappa = 0.1$, $0.3$, $0.5$; $68\%$
    probability region for $\gamma$ obtained using the combined method
    (filled area) or the GL method (horizontal lines) as a function of
    $\kappa$.}
  \label{fig:gammafull}
\end{figure}

\begin{figure}[t]
  \centering
  \includegraphics[width=.25\textwidth]{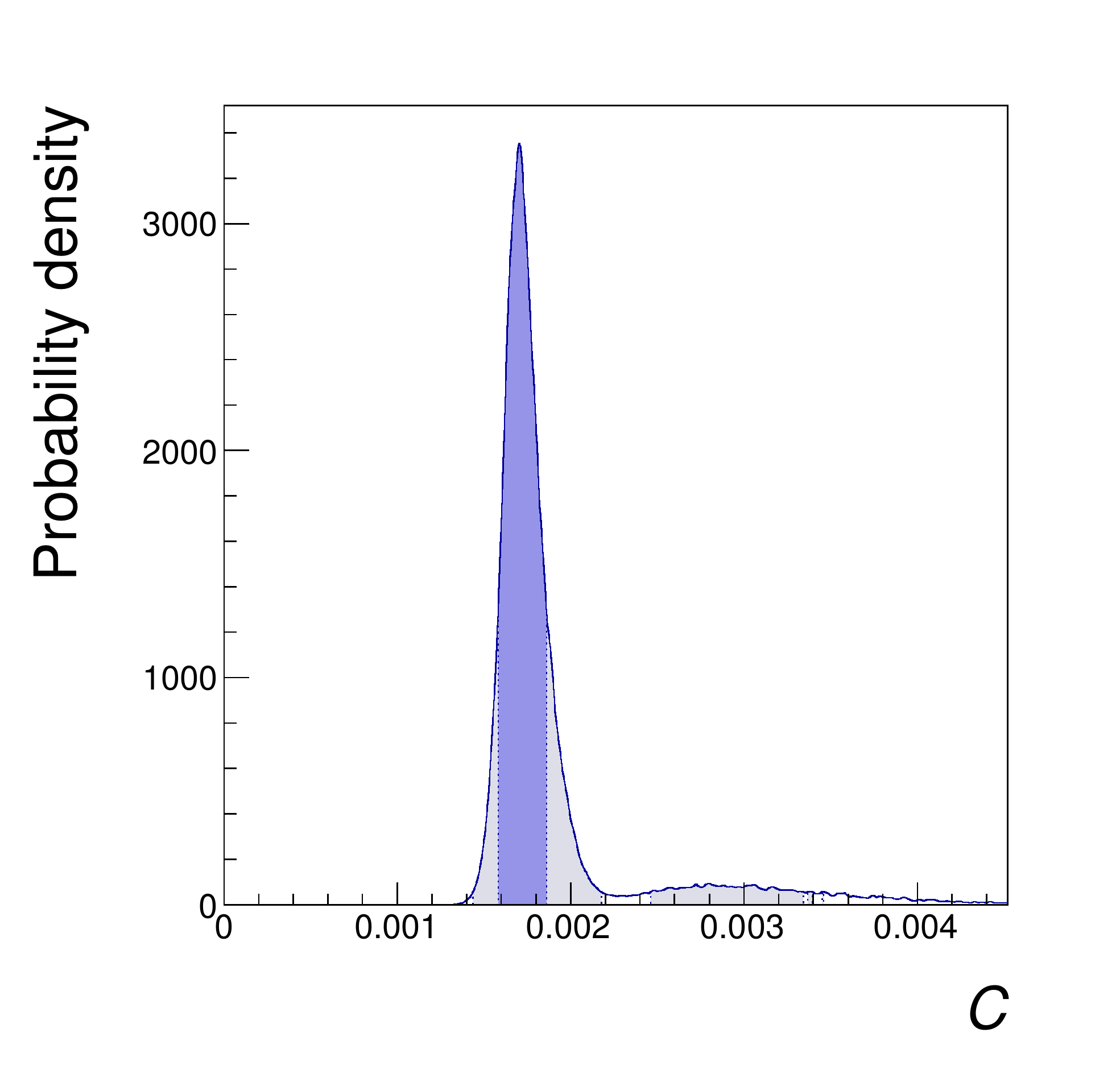}
  \includegraphics[width=.25\textwidth]{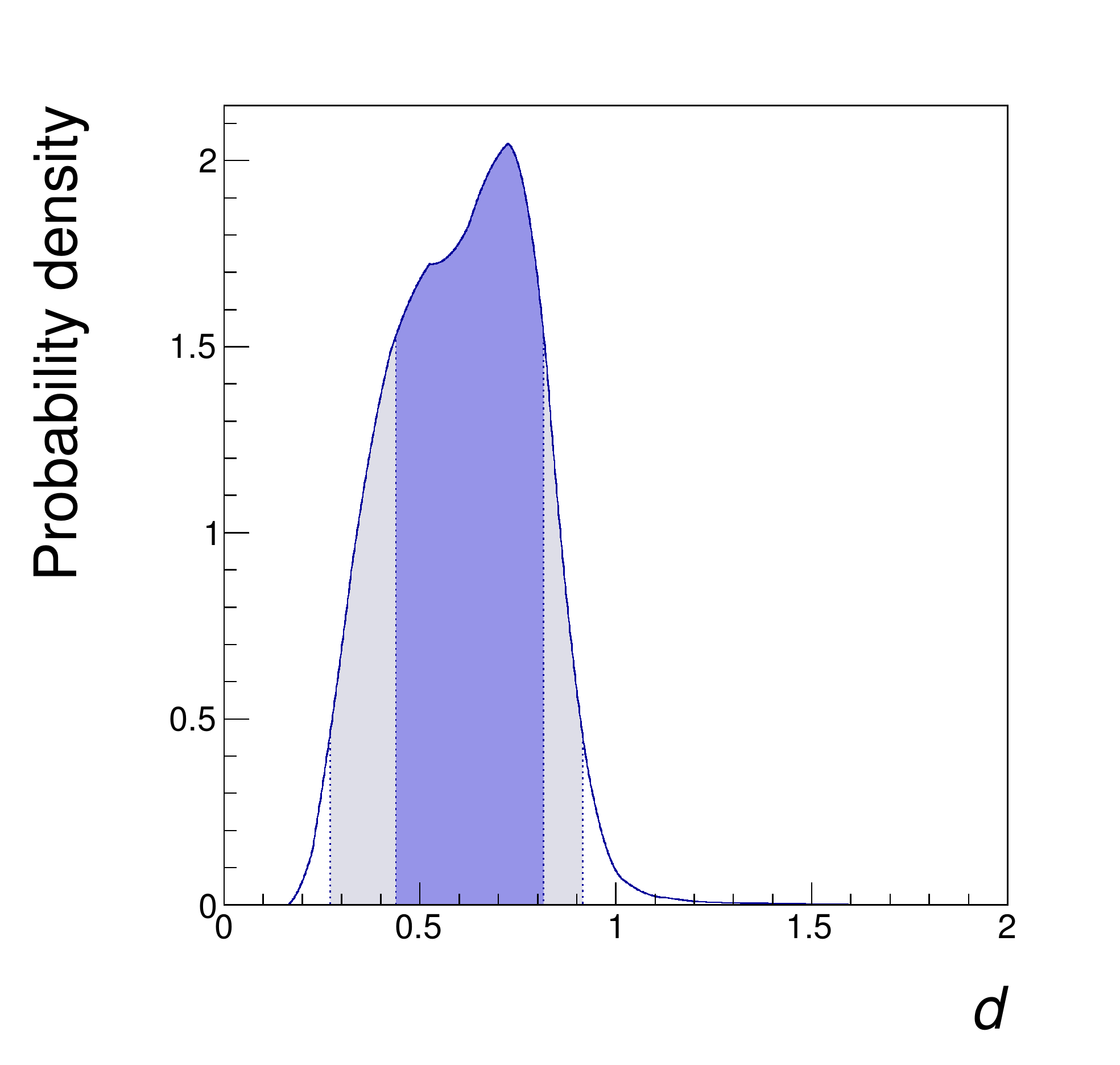}
  \includegraphics[width=.25\textwidth]{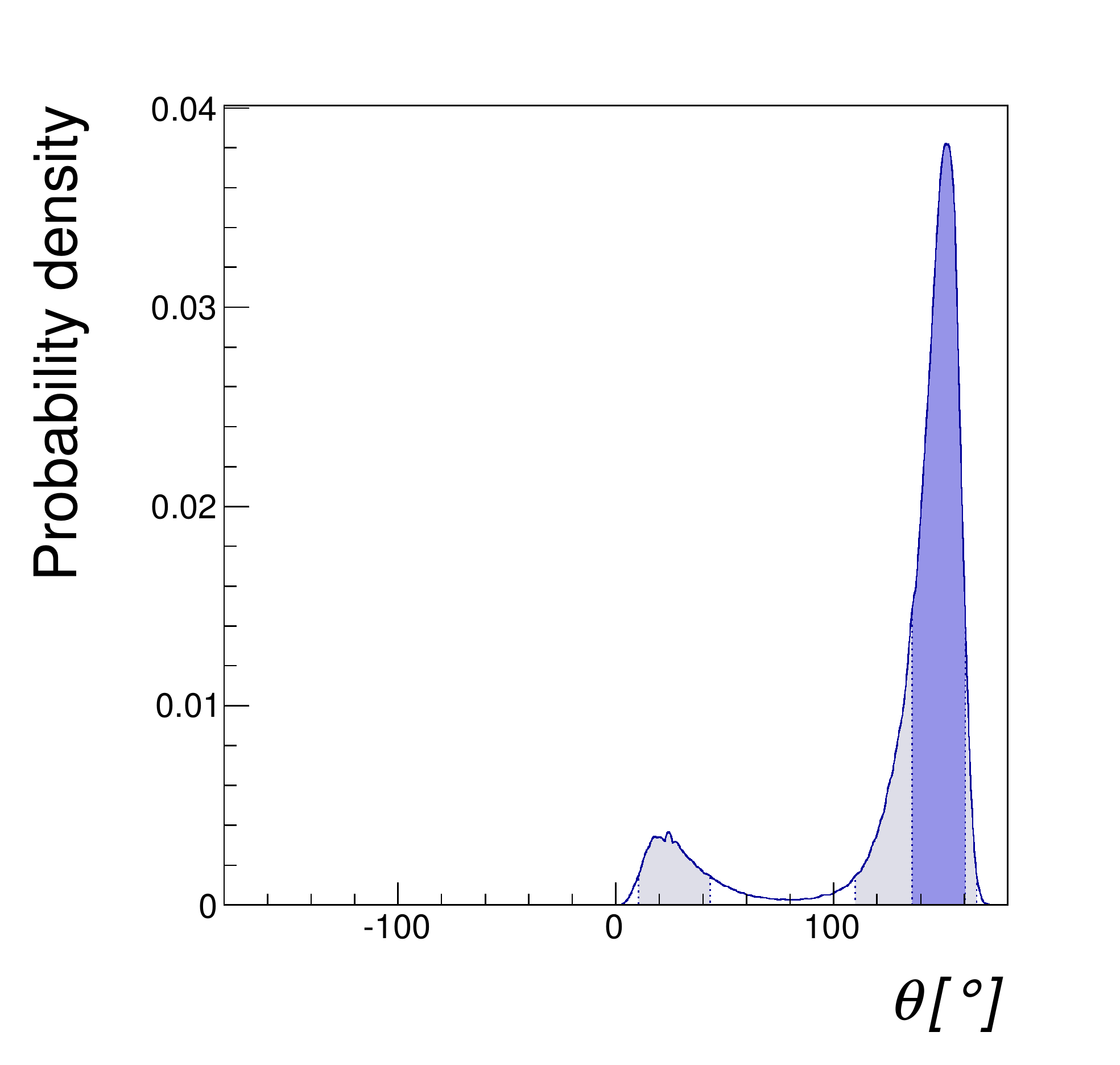}
  \includegraphics[width=.25\textwidth]{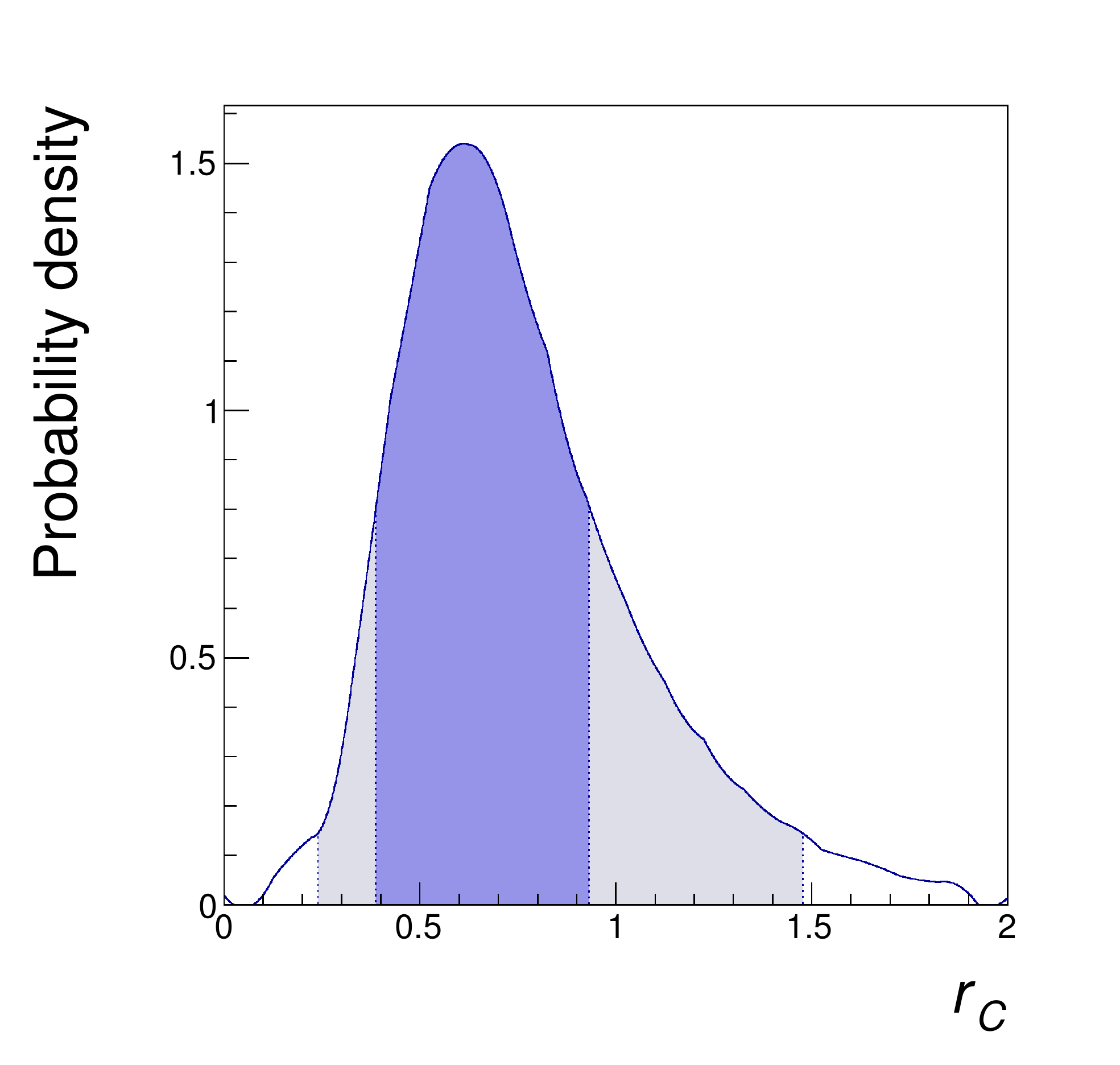}
  \includegraphics[width=.25\textwidth]{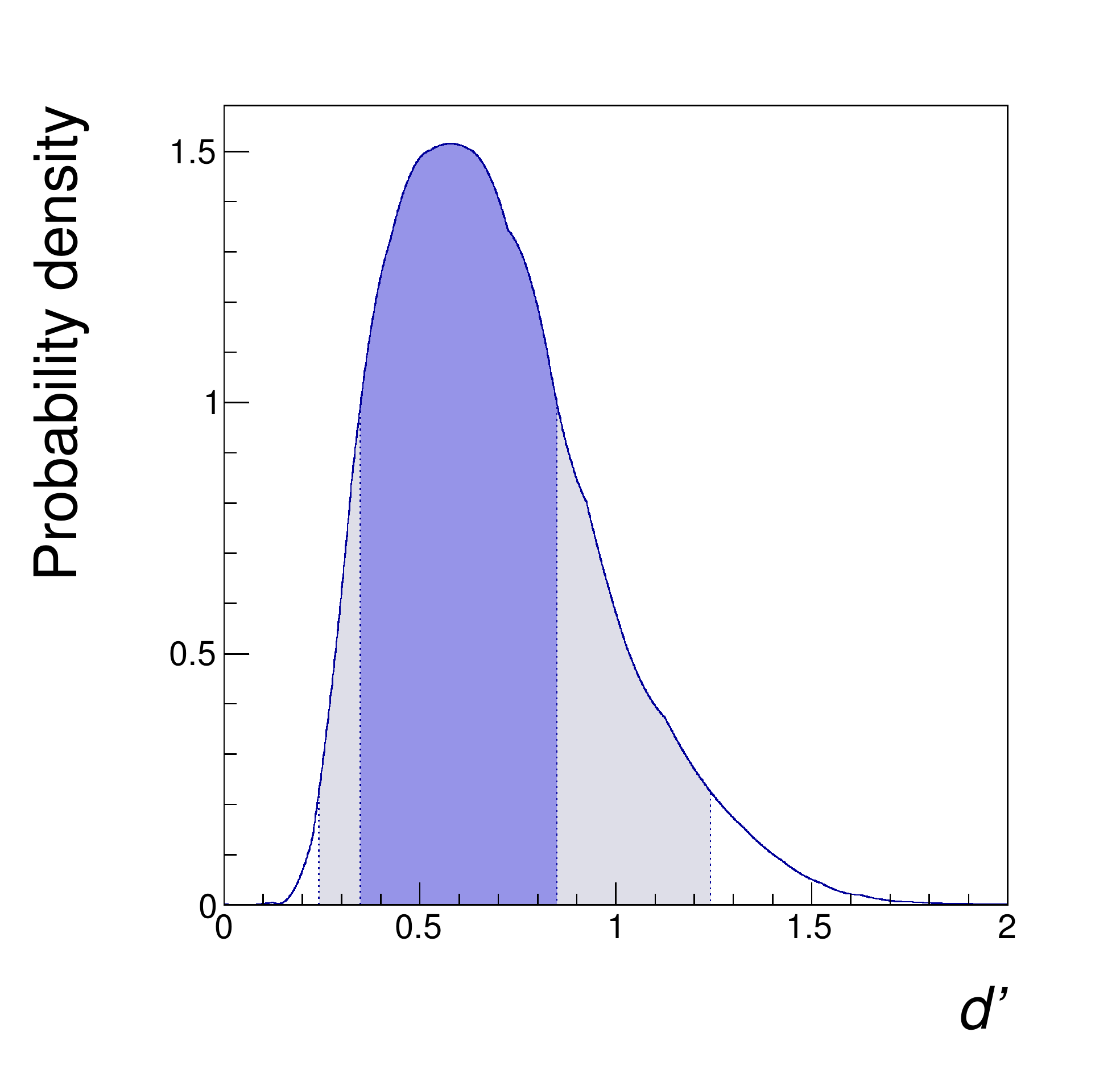}
  \includegraphics[width=.25\textwidth]{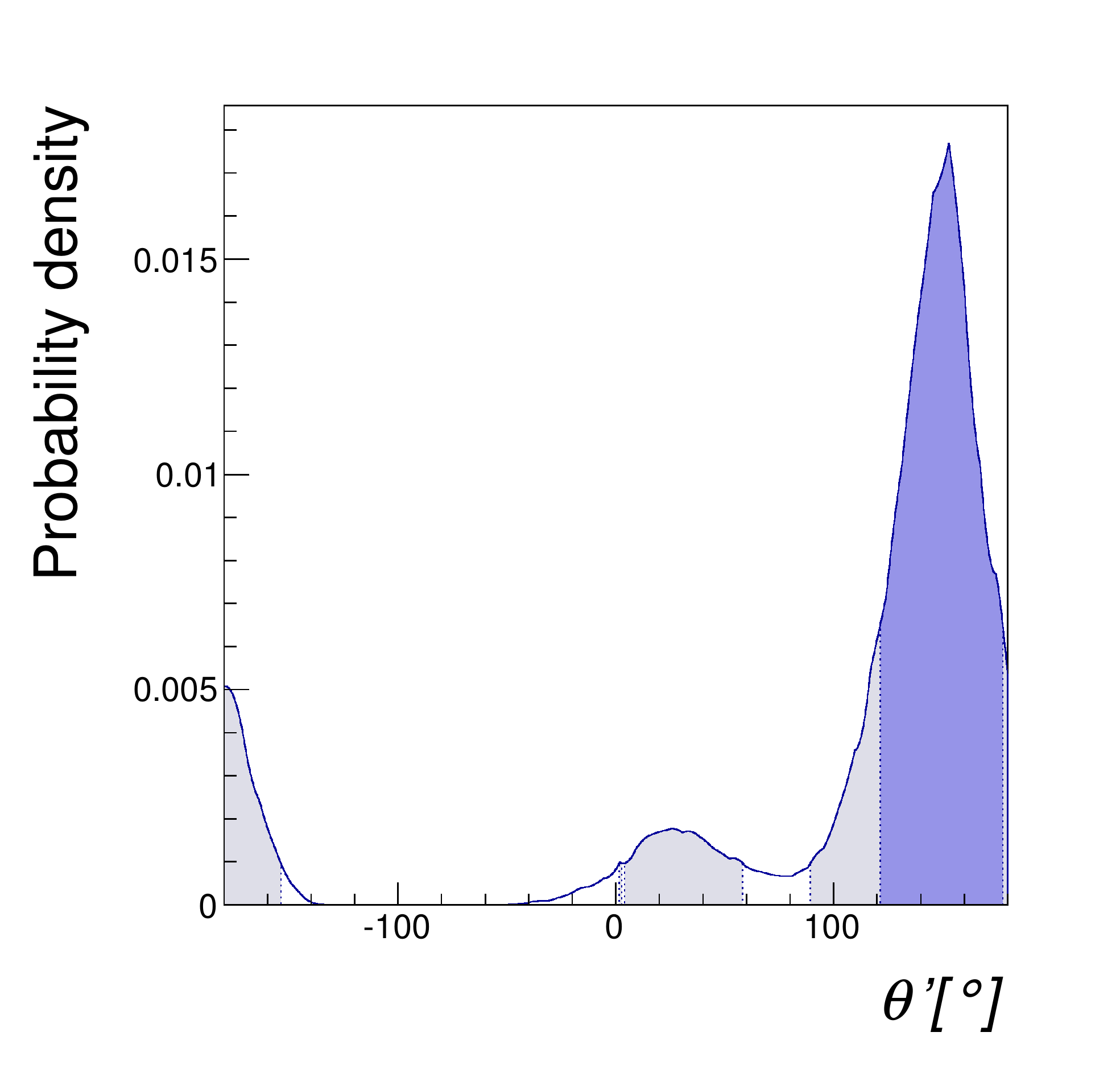}
  \caption{From left to right and from top to bottom: P.d.f. for $C$,
    $d$, $\theta$, $r_C$, $d^\prime$, $\theta^\prime$ obtained using the
    combined method for $\kappa = 0.9$.}
  \label{fig:uspinb}
\end{figure}

We now consider the result of the combined GL+F analysis. In
Fig.~\ref{fig:gammafull} we present the p.d.f. for $\gamma$ for
$\kappa = 0.1$, $0.3$ and $0.5$. We see that the result of the
combined analysis is much more stable against the amount of U-spin
breaking allowed. We also plot the $68\%$ probability region for
$\gamma$ obtained using the combined method as a function of $\kappa$,
and compare it to the GL result. We see that there is a considerable
gain in precision even for gigantic values of $\kappa$. Actually, as
can be seen in Fig.~\ref{fig:uspinb}, where the posteriors for
hadronic parameters and the U-spin breaking parameter $r_C$ are
reported, the $68\%$ probability range for $r_C$ is between $\sim 0.4$ and
$\sim 0.9$. The fact that the $r_C$ posterior is not centered around $1$,
but the product $r_C r_\mathrm{fact}$ is close to $1$, may signal a
failure of factorization and/or of the QCD sum rule estimate of
$r_\mathrm{fact}$. On the other hand, the posteriors for $d^\prime$ and
$\theta^\prime$ are well compatible with small U-spin breaking. In any
case, we think that the lesson to be learned from
Fig.~\ref{fig:uspinb} is that values of $\kappa$ up to $0.6$ or $0.7$
cannot be excluded, but nevertheless the combined method remains
useful. This happens because the peak around $\gamma \sim 30^\circ$ in
the GL result corresponds to values of $\theta$ that are different
from the ones needed in the F analysis to obtain similar values of
$\gamma$, while the peak at $\gamma \sim 70^\circ$ is obtained for the
same values of hadronic parameters in both the GL and F analyses.

\begin{figure}[t]
  \centering
  \includegraphics[width=.23\textwidth]{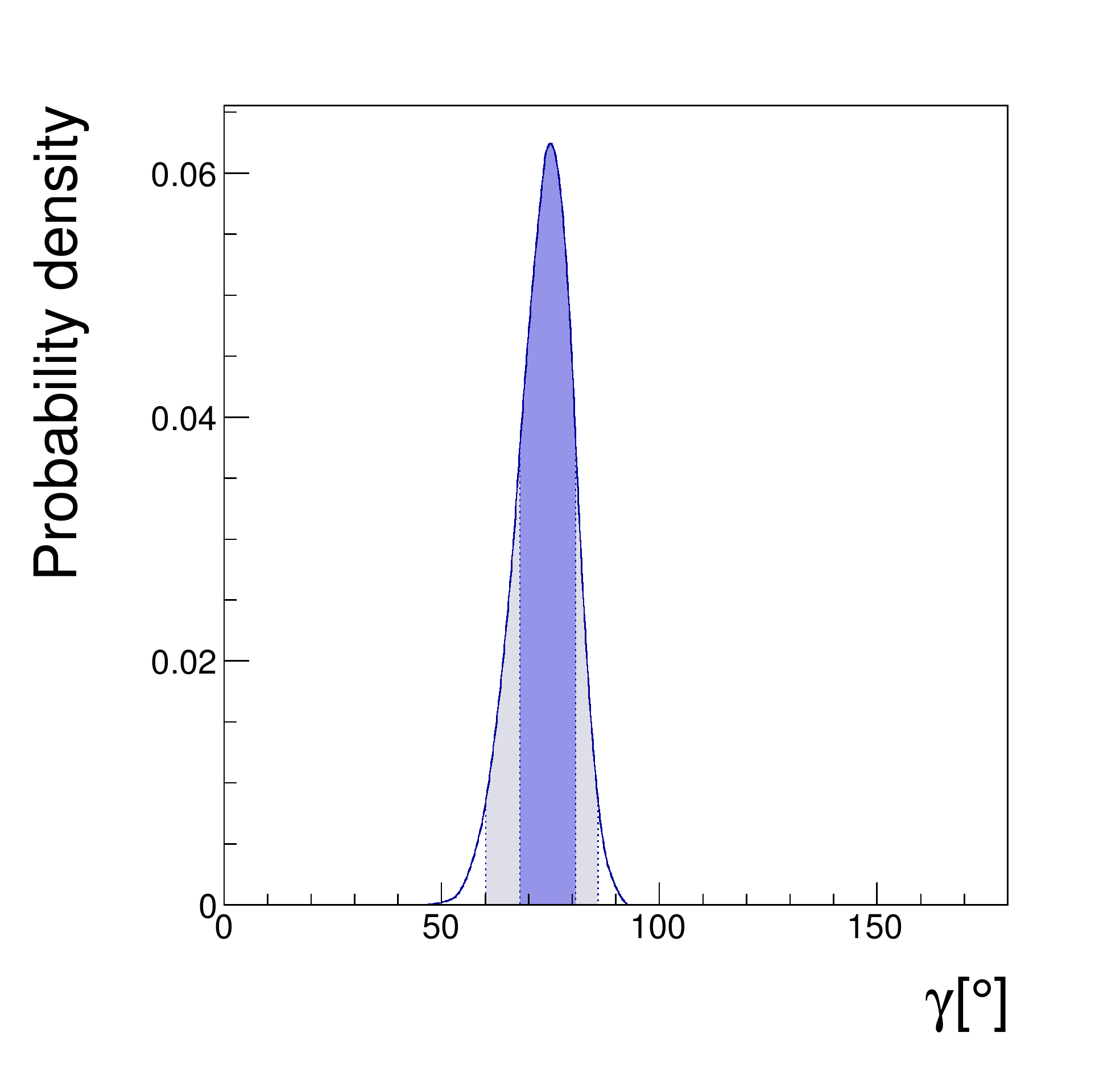}
  \includegraphics[width=.23\textwidth]{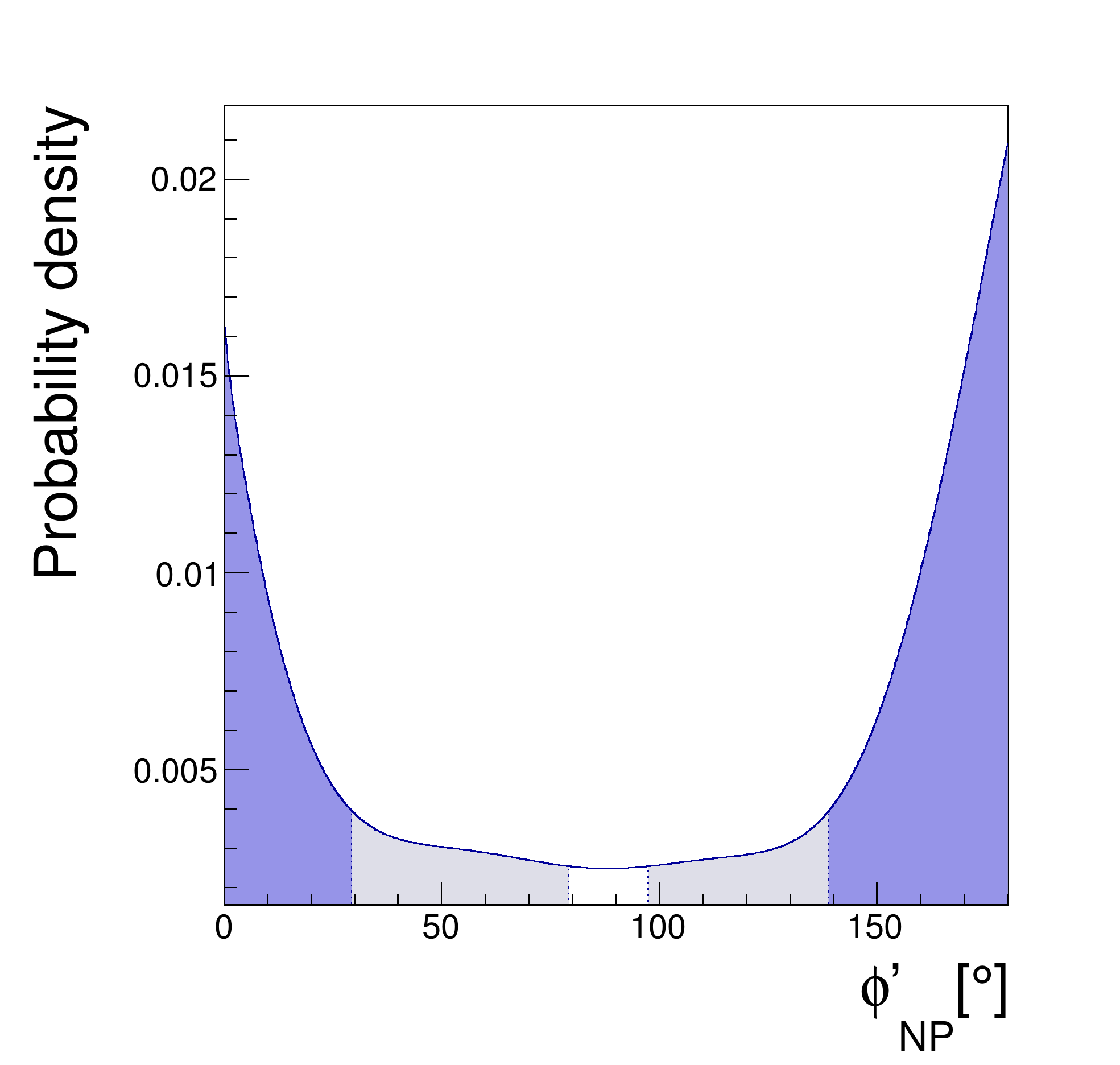}
  \includegraphics[width=.23\textwidth]{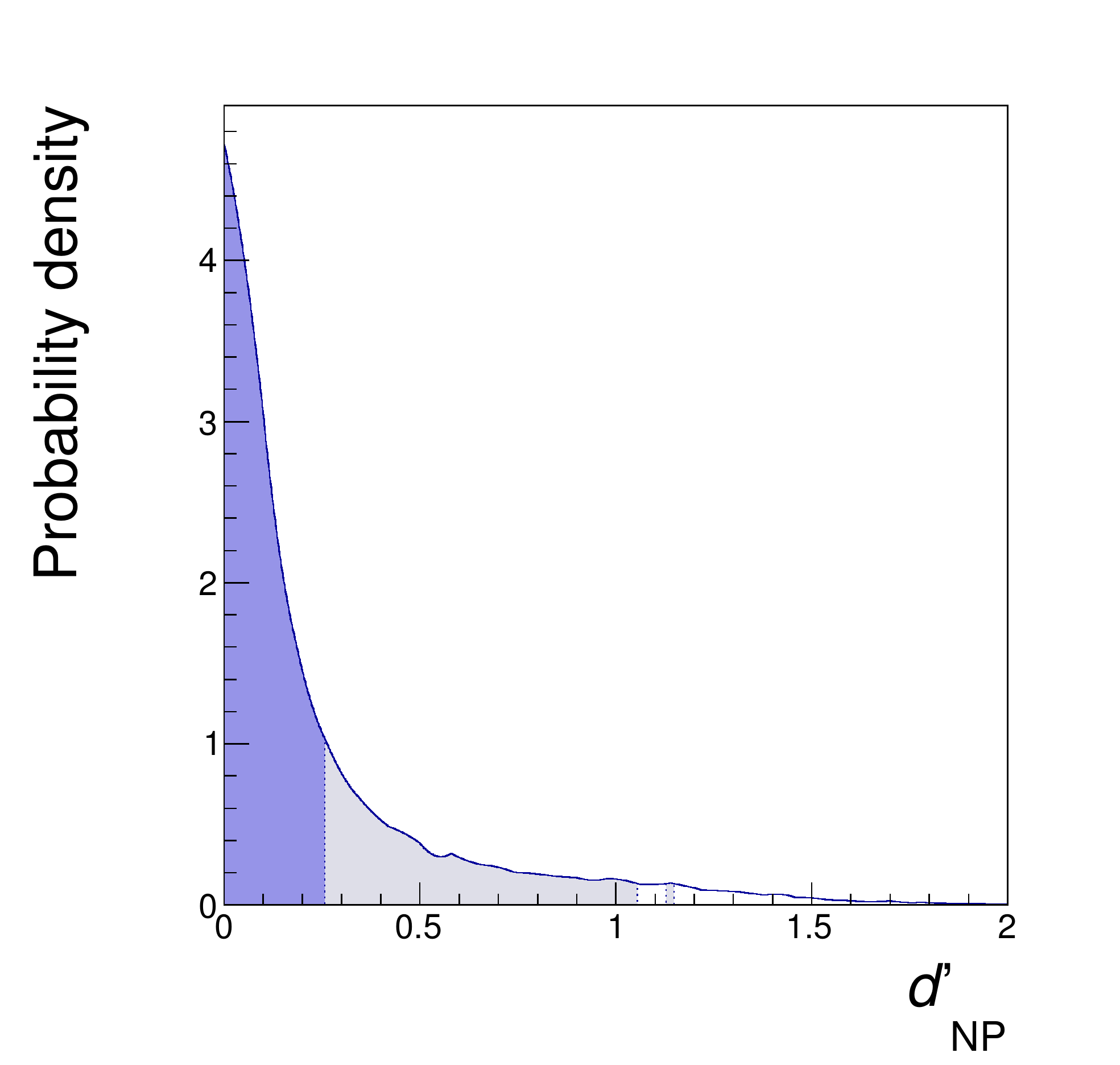}
  \includegraphics[width=.23\textwidth]{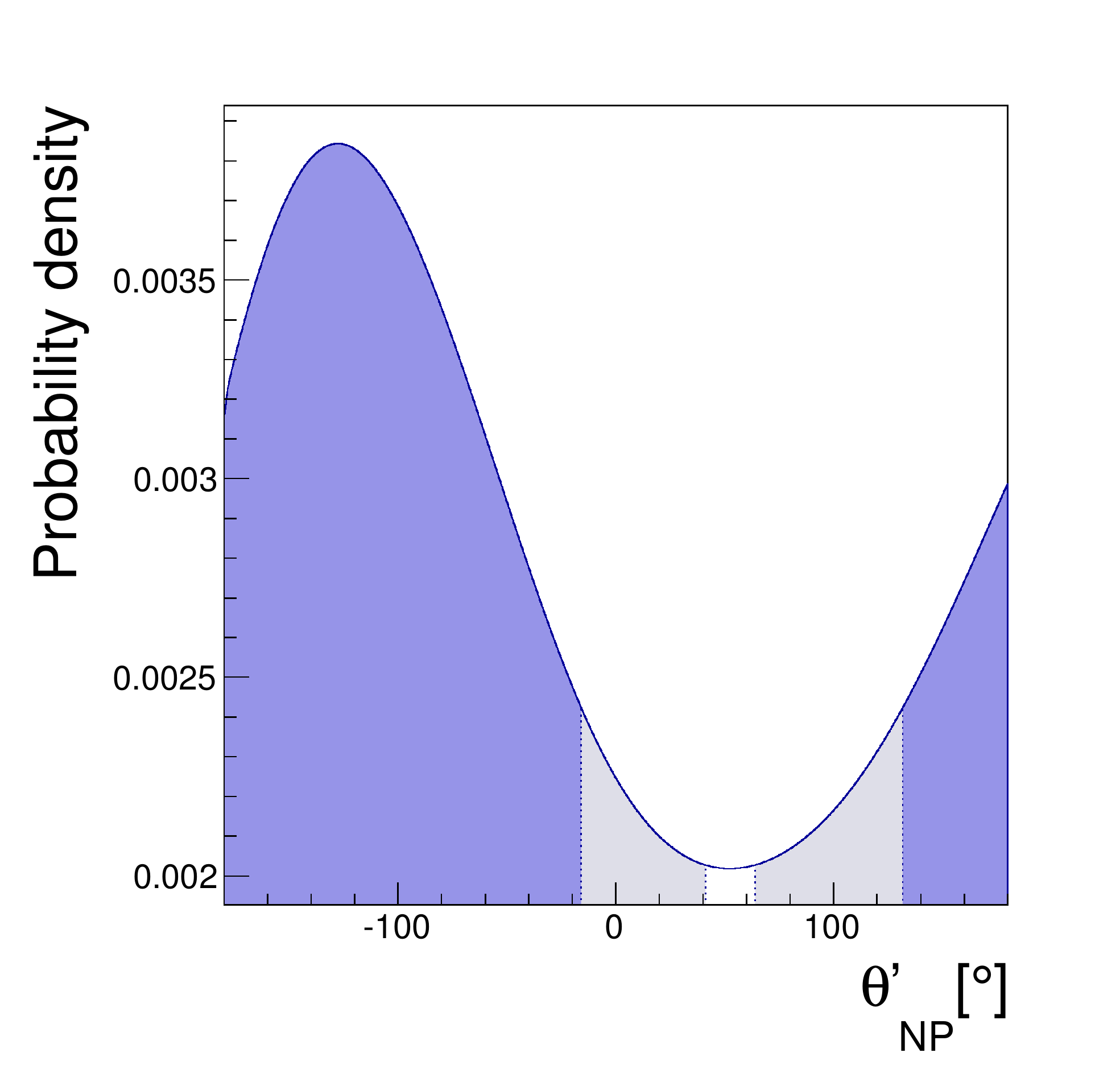}
  \caption{From left to right: P.d.f. for $\gamma$,
    $\phi_\mathrm{NP}^\prime$, $d^\prime_\mathrm{NP}$ and
    $\theta^\prime_\mathrm{NP}$ obtained  using the
    combined method for $\kappa = 0.5$.}
  \label{fig:np}
\end{figure}

New Physics could affect the determination of $\gamma$ in the combined
method by giving (electroweak) penguin contributions with a new
CP-violating phase. Let us assume for concreteness that NP only enters
$b \to s$ decays, so that the isospin analysis of the GL channels is
still valid. In the framework of a global fit, one can simultaneously
determine $\gamma$ and the NP contribution to $b \to s$ penguins. For
the purpose of illustration, we can just use as input the value of
$\gamma$ from tree-level processes, $\gamma_\mathrm{tree} = (76 \pm
9)^\circ$ \cite{*[][{ and online update at
    http://www.utfit.org/UTfit/ResultsSummer2011PostLP}] Bona:2007vi},
and look at the posterior for $\gamma$ and for the NP penguin
amplitude. Taking for simplicity equal magnitudes for the
  NP contribution to $A(B_s \to K^+ K^-)$ and $A(\bar B_s \to K^+
  K^-)$, we write
\begin{eqnarray}
  \label{eq:npampli}
  A(B_s \to K^+ K^-) &=& C^\prime \frac{\lambda}{1-\lambda^2/2} 
  (e^{i \gamma} + \frac{1-\lambda^2}{\lambda^2} (d^\prime e^{i
    \theta^\prime} + e^{i 
    \phi_\mathrm{NP}^\prime} d^{\prime}_\mathrm{NP} e^{i
    \theta^\prime_\mathrm{NP}}))\,, \\
  A(\bar B_s \to K^+ K^-) &=& C^\prime \frac{\lambda}{1-\lambda^2/2}
   (e^{-i \gamma} + \frac{1-\lambda^2}{\lambda^2} (d^\prime e^{i
     \theta^\prime} +  e^{-i
    \phi_\mathrm{NP}^\prime} d^\prime_\mathrm{NP} e^{i
    \theta^\prime_\mathrm{NP}}))\,.\nonumber
\end{eqnarray}
Taking uniformly distributed $d^\prime_\mathrm{NP} \in [0,2]$ and
$\phi^\prime_\mathrm{NP}, \theta^\prime_\mathrm{NP}\in [-\pi,\pi]$ we obtain
the p.d.f. reported in Fig.~\ref{fig:np} for $\kappa =
0.5$. It yields $\gamma = (74 \pm 6)^\circ$, and a $95\%$
probability upper bound on $d^\prime_\mathrm{NP}$ around $1$. The
bound is actually much stronger for large values of
$\phi^\prime_\mathrm{NP}$.

Finally, we notice that $B_s \to K K$ decays can also be used to
obtain information on $\phi_M(B_s)$. The optimal choice in this
respect is represented by $B_s \to K^{(*)0} \bar K^{(*)0}$ (with $B_d
\to K^{(*)0} \bar K^{(*)0}$ as U-spin related control channel to
constrain subleading contributions), since in this channel there is no
tree contribution proportional to $e^{i \gamma}$
\cite{Ciuchini:2007hx}. However, the combined analysis described
above, in the framework of a global SM fit, can serve for the same
purpose. To illustrate this point, we perform the GL+F analysis not
using the measurement of $2 \beta_s$ from $b \to c \bar c s$
decays. In this way, we obtain $2 \beta_s = (6 \pm 14)^\circ$ for
$\kappa = 0.5$. With improved experimental accuracy, this
determination will become competitive with the one from $b \to c \bar
c s$ decays, since the theoretical uncertainty can be estimated more
reliably in the case of $B_s \to K^+ K^-$ decays, waiting for
time-dependent analyses of the $B_{(s)} \to K^{(*)0} \bar K^{(*)0}$
channels.\footnote{The proposal of ref.~\cite{Ciuchini:2007hx} has
  been recently critically reexamined in
  ref.~\cite{Bhattacharya:2012hh}. We notice that the present analysis
  shows no particular enhancement of the contribution proportional to
  $e^{i \gamma}$ in $B_s \to K^+ K^-$, in agreement with the
  expectation that $B_{s} \to K^{(*)0} \bar K^{(*)0}$ should be
  penguin-dominated to a very good accuracy.} To illustrate the
potential of this method, we have repeated the analysis reducing the
experimental uncertainty on $A_\mathrm{CP}(B_d \to \pi^+\pi^-)$,
$S(B_d \to \pi^+\pi^-)$, $A_\mathrm{CP}(B_s \to K^+K^-)$ and $S(B_s
\to K^+K^-)$ down to $\pm 0.02$. With such small experimental errors,
it becomes crucial to take correctly into account the effect of the
subleading term proportional to $e^{i\gamma}$ in the amplitude (this is the case for any channel
used to extract $\beta_s$ with an uncertainty of few degrees, including
$B_s\to J/\psi\phi$). This is best done in the context of a global fit.
For the purpose of illustration, we take as input the SM fit result
$\gamma = (69.7 \pm 3.1)^\circ$ \cite{*[][{ and online update at
    http://www.utfit.org/UTfit/ResultsSummer2011PostLP}] Bona:2007vi} and
obtain $2 \beta_s = (2.6 \pm 2.7)^\circ$ for $\kappa =
0.5$. The error, which includes the theoretical uncertainty, could be
further reduced improving the other relevant measurements, including
the $B_d$ decay modes, and by adding the $B_{d,s} \to K^{(*)0} \bar
K^{(*)0}$ channels, allowing to test the SM prediction for CP
violation in $B_s$ mixing.

To conclude, let us summarize our findings. We suggest that the usual
GL analysis to extract $\alpha$ from $B_d \to \pi\pi$ be supplemented
with the inclusion of the $B_s \to K^+ K^-$ modes, in the framework of
a global CKM fit. The method optimizes the constraining power of these
decays and allows to derive constraints on NP contributions to penguin
amplitudes or on the $B_s$ mixing phase. We have illustrated these
capabilities with a simplified analysis, neglecting correlations with
other SM observables.

M.C. is associated to the Dipartimento di Fisica, Universit\`a di Roma
Tre. E.F., S.M. and L.S. are associated to the Dipartimento di Fisica,
Universit\`a di Roma ``La Sapienza''. We acknowledge partial support
from ERC Ideas Starting Grant n.~279972 ``NPFlavour'' and ERC Ideas
Advanced Grant n.~267985 ``DaMeSyFla''. We are indebted to Vincenzo
Vagnoni for enlightening comments and suggestions and for pointing out
to us the correction factor derived in ref.~\cite{deBruyn:2012wj}.

\bibliography{bskpkm}
 
\end{document}